\documentclass[10pt,aps,prd,twocolumn,preprintnumbers,superscriptaddress,amsmath,amssymb,nofootinbib]{revtex4-1}

\usepackage[hidelinks]{hyperref}
\hypersetup{colorlinks, linkcolor={blue}, citecolor={blue}, urlcolor={blue}}
\usepackage[font=small, labelfont=bf, justification = raggedright]{caption}
\usepackage{graphicx}
\graphicspath{ {./Figures/} }
\let\subcaption\relax

\usepackage{subcaption}
\usepackage{bm}

\footnotetext{These authors contributed equally to this work.}

\begin{document}
\preprint{Prepared for submission to NeurIPS}
\title{\texttt{ODE}$\mathcal{N}$: A Framework to Solve Ordinary \\Differential Equations using Artificial Neural Networks}

\author{Liam L. H. Lau$^\mathsection \,$}
\email[]{lhll2@cam.ac.uk}
\affiliation{Gonville \& Caius College, Trinity Street, Cambridge, CB2 1TA, UK}

\author{Denis Werth$^\mathsection \,$}
\email[]{denis.werth@ens-paris-saclay.fr}
\affiliation{\'Ecole Normale Sup\'erieure Paris-Saclay, Department of Physics, France}

\begin{center}
 \begin{abstract}
We explore in detail a method to solve ordinary differential equations using feedforward neural networks. We prove a specific loss function, which does not require knowledge of the exact solution, to be a suitable standard metric to evaluate neural networks' performance. Neural networks are shown to be proficient at approximating continuous solutions within their training domains. We illustrate neural networks' ability to outperform traditional standard numerical techniques. Training is thoroughly examined and three universal phases are found: (i) a prior tangent adjustment, (ii) a curvature fitting, and (iii) a fine-tuning stage. The main limitation of the method is the nontrivial task of finding the appropriate neural network architecture and the choice of neural network hyperparameters for efficient optimization. However, we observe an optimal architecture that matches the complexity of the differential equation. A user-friendly and adaptable open-source code (\texttt{ODE}$\mathcal{N}$) is provided on GitHub.

\end{abstract}   
\end{center}

\maketitle

\section{Introduction}
\label{sec:introduction}

Neural networks (NNs) are known to be powerful tools due to their role as universal continuous function approximators \cite{Hornik}. Rewarding NNs for their improving performances in specific tasks can be done by minimizing a specially designed loss function. In fact, the last ten years have seen a surge in the capabilities of NNs, specifically deep neural networks, due to the increase in supervised data and compute. For example, one can cite AlexNet \cite{AlexNet}, a convolutional neural network for image recognition and GPT-2 \cite{gpt2}, a powerful Transformer language model. 

Although NNs and machine learning are used in physics \cite{Carleo, Tanaka, Li_Tan_Jiang, Wetzel, Wei, Huang, Caldeira, Shen, Arai, Yoshioka}, the use of NNs has still not been widely accepted in the scientific community due to their black-box nature, the unclear and often unexplored effects of hyperparameter choice for a particular problem, and a large computational cost for training. 

Usually, neural networks are used for problems involving dimensional reduction \cite{Wang_2014_CVPR_Workshops}, data visualization, clustering \cite{gene}, and classification \cite{drug}. Here, we use the potential of NNs to solve differential equations. Differential equations are prevalent in many disciplines including Physics, Chemistry, Biology, Economics, and Engineering. When an analytical solution cannot be found, numerical methods are employed and are very successful \cite{breuer,kang,yan}. However, some systems exhibit differential equations that are not efficiently solved by usual numerical methods. Such differential equations can be numerically solved to the required accuracy by novel specific codes, such as \textsf{oscode} \cite{Agocs} which efficiently solves one-dimensional, second-order, ordinary differential equations (ODEs) with rapidly oscillating solutions which appear in cosmology and condensed matter physics \cite{wensch} among other systems. Yet, specifically written algorithms \cite{beyond} and codes require a lot of time and resources.\\
\indent{}In this paper, we ask whether the unsupervised method of solving ODEs using NNs, first developed by Lagaris et al. \cite{Lagaris} and improved in \cite{Piscopo, Koryagin, Liu, lu, Mall, Michoski, irina, sirignano}, is robust. In previous investigations, there has been no clear consensus on what exact method should be used. Multiple loss functions were used, sometimes requiring the exact solution to be known \cite{bekele} \cite{Raissi2017physicsID}, and the boundary/initial conditions were either included in the model as an extra term in the loss function or treated independently by means of trial functions. No explicit benchmark to evaluate the NN performances was transparently introduced. Training and extrapolation performances were not meticulously studied, making NNs appear as black-boxes.\\
\indent{}This work aims at filling the mentioned gaps and demystifying a few aspects. We provide a clear step-by-step method to solve ODEs using NNs. We prove a specific loss function from the literature \cite{Piscopo} to be an appropriate benchmark metric to evaluate the NN performances, as it does not require the exact solution to be known. The training process is investigated in detail and is found to exhibit three phases that we believe are the general method that the NN learns to solve a given ODE with boundary/initial conditions. We explore the effect of training domain sampling, extrapolation performances and NN architectures. Contrary to numerical integration, approximating the ODE solution by a NN gives a function that can be continuously evaluated over the entire training domain. However, finding an appropriate NN architecture to reach the desired accuracy requires extensive testing and intuition. An open-source and user-friendly code (\texttt{ODE}$\mathcal{N}$) that accompanies this manuscript\footnote{\href{https://github.com/deniswerth/ODEN}{https://github.com/deniswerth/ODEN}} is provided, along with animations that help to visualize the training process.

In the next section, we briefly present the basics of NNs, followed by the method used to solve ODEs with NNs in Section \ref{sec:methodology}. In Section \ref{sec:applications}, three different ODEs are solved as illustration. A specific loss function is proved to be a universal metric to assess the accuracy and different simple NN models are tested in Section \ref{sec:perfomances}, particularly highlighting the role of complexity when choosing the NN. Finally, conclusion and outlook are provided in Sections \ref{sec:conclusion}.


\section{Background}
\label{sec:background}

The basic unit of a NN is a neuron $i$ that takes a vector of input features $\bm{x} = (x_1, x_2, ...)$ and produces a scalar output $a_i(\bm{x})$. In almost all cases, $a_i$ can be decomposed into a linear operation taking the form of a dot product with a set of weights $\bm{w}^{(i)} = (w_1^{(i)}, w_2^{(i)}, ...)$ followed by re-centering with an offset called the bias $b^{(i)}$, and a non-linear transformation \textit{i.e.} an activation function $\sigma_i : \mathbb{R}\rightarrow [0,1]$ which is usually the same for all neurons. One can write the full input-output formula for one neuron as follows

\begin{equation}
a_i(\bm{x}) = \sigma_i\left( \bm{w}^{(i)}.\,\bm{x} + b^{(i)}  \right).
\end{equation}

A NN consists of many such neurons stacked into layers, with output of one layer serving as input for the next. Thus, the whole NN can be thought of as a complicated non-linear transformation of inputs $x$ into an output $\mathcal{N}$ that depends on the weights and biases $\bm{\theta}$ of all the neurons in the input, hidden, and output layers \cite{bishop2007}:

\begin{equation}
\text{Neural Network} = \mathcal{N}(x, \bm{\theta}).
\end{equation}

The NN is trained by finding the value of $\bm{\theta}$ that minimizes the loss function $\mathcal{L}(\bm{\theta})$, a function that judges how well the model performs on corresponding the NN inputs to the NN output. Minimization is usually done using a gradient descent method \cite{bishop2007}. These methods iteratively adjust the NN parameters $\bm{\theta}$ in the direction (on the parameter surface) where the gradient $\mathcal{L}$ is large and negative. In this way, the training procedure ensures the parameters $\bm{\theta}$ flow towards a local minimum of the loss function. The main advantage of NNs is to exploit their layered structure to compute the gradients of $\mathcal{L}$ in a very efficient way using backpropagation \cite{bishop2007}.


\section{Method}
\label{sec:methodology}

The starting point of solving ODEs using NNs is the universal approximation theorem. The theorem states that a feedforward NN with hidden layers containing a finite number of neurons can approximate any continuous functions at any level of accuracy \cite{Hornik}. Thus, one may expect NNs to perform well on solving ODEs. Despite the existence of such NNs, the theorem does not provide us with a recipe to find these NNs nor touch upon their learnabilities.

Differential equations with an unknown solution $f$ can be written as a function $\mathcal{F}$ of $f$ and their derivatives in the form

\begin{equation}
\label{diff_eq}
\mathcal{F}[x, f(x), \nabla f(x), ... ,\nabla^{p}f(x)] = 0 \hspace*{0.3cm}\text{for}\hspace*{0.3cm} x\in \mathcal{D},
\end{equation}
with some Dirichlet/Neumann boundary conditions or initial conditions, where $\mathcal{D}$ is the differential equation domain and $\nabla^{p}f(x)$ is the $p$-th gradient of $f$. The notations and the approach can easily be generalized to coupled and partial differential equations. Following classic procedures, we then discretize the domain $\mathcal{D}$ in $N$ points $\bm{x} = (x_1, x_2, ..., x_N)$. For each point $x_i\in \mathcal{D}$, Eq. (\ref{diff_eq}) must hold so that we have

\begin{equation}
\mathcal{F}[x_i, f(x_i), \nabla f(x_i), ..., \nabla^{p}f(x_i)] = 0.
\end{equation}

Writing the differential equation in such a way allows us to convert the problem of finding solutions to an optimization problem. Indeed, an approximated solution is a function that minimizes the square of the left-hand side of Eq. (\ref{diff_eq}). In our approach, we identify the approximated solution of the differential equation (\ref{diff_eq}) with a NN $\mathcal{N}(x, \bm{\theta})$ and express the loss function as follows \cite{Piscopo}:

\begin{equation}
\label{cost_function}
\begin{aligned}
\mathcal{L}(\bm{\theta}) &= \frac{1}{N}\sum_{i=1}^N \mathcal{F}\left[x_i, \mathcal{N}(x_i, \bm{\theta}), \nabla \mathcal{N}(x_i, \bm{\theta}), ... , \nabla^{p}\mathcal{N}(x_i, \bm{\theta})\right]^2 \\
&+ \sum_{j} [\nabla^{k}\mathcal{N}(x_{j}, \bm{\theta}) - K_{j}]^2,
\end{aligned}
\end{equation} 
with $K_{j}$ being some chosen constants. The loss function contains two terms: the first one forces the differential equation (\ref{diff_eq}) to hold, and the second one encodes the boundary/initial conditions\footnote{Usually, $k = 0$ and/or $k=1$.}. Note that $\mathcal{L}(\bm{\theta})$ does not depend on the exact solution\footnote{In some works, the mean squared error \cite{bekele} \cite{Raissi2017physicsID} is used.}. The approximated solution $\mathcal{N}(x, \bm{\theta})$ is found by finding the set of weights and biases $\bm{\theta}$ such that the loss function $\mathcal{L}(\bm{\theta})$ is minimized on the training points $x_i$ for $i = 1, 2, ..., N$. Solving ODEs using NNs can then be expressed in terms of the following steps:

\begin{enumerate}
	\item Write the differential equation in the form shown in Eq. (\ref{diff_eq});
	\item Discretize the differential equation domain in $N$ points $\bm{x} = (x_1, x_2, ..., x_N)$;
	\item Construct a neural network $\mathcal{N}(x, \bm{\theta})$ by choosing its architecture \textit{i.e.} the number of hidden layers and the number of neurons in each layer, and the activation functions;
	\item Write the loss function $\mathcal{L}(\bm{\theta})$ in the form shown in Eq. (\ref{cost_function}) including boundary/initial conditions;
	\item Minimize $\mathcal{L}(\bm{\theta})$ to find the optimal NN parameters $\bm{\theta}^{\star}$ \textit{i.e.} train the NN;
	\item Once the NN has been trained on the discrete domain, $\mathcal{N}(x, \bm{\theta}^{\star})$ is an approximation of the exact solution that can be evaluated continuously within the training domain.
\end{enumerate}

For our setup, as ODE solutions are scalar functions of a scalar independent variable, the NN has one input neuron and one output neuron. The first term of $\mathcal{L}(\bm{\theta})$ in Eq. (\ref{cost_function}) requires computing the gradients of the NN output with respect to the NN input. Gradient computation, as well as the network implementation and the training procedure, is performed using the \textsf{Keras} \cite{keras} framework with a \textsf{Tensorflow 2} \cite{tensorflow} backend. The general method can be easily extended to coupled and partial differential equations\footnote{For example, coupled differential equations can be solved by corresponding the number of output neurons to the number of equations in the system one wants to solve.}.


\section{Applications}
\label{sec:applications}
    
The method is verified against the exact solution of first and second order ODEs. We then use the method to solve an ODE that exhibits rapid oscillations. Training processes were computed using 12 CPU cores and took between a few seconds to a few minutes, depending on the number of epochs and the NN complexity.

    \subsection{First order ordinary differential equation}
    
We first demonstrate the accuracy and efficiency of the NN solver when applied to the following simple first order differential equation with $x \in [0,2]$, subject to the initial value $f(0) = 0$:

\begin{equation}
\label{first_order}
\frac{\mathrm{d}f}{\mathrm{d}x}(x) + f(x) = e^{-x} \cos(x).
\end{equation}

\begin{figure}[h!]
	\centering
	\hspace*{-0.5cm}
	\includegraphics[width=0.5\textwidth]{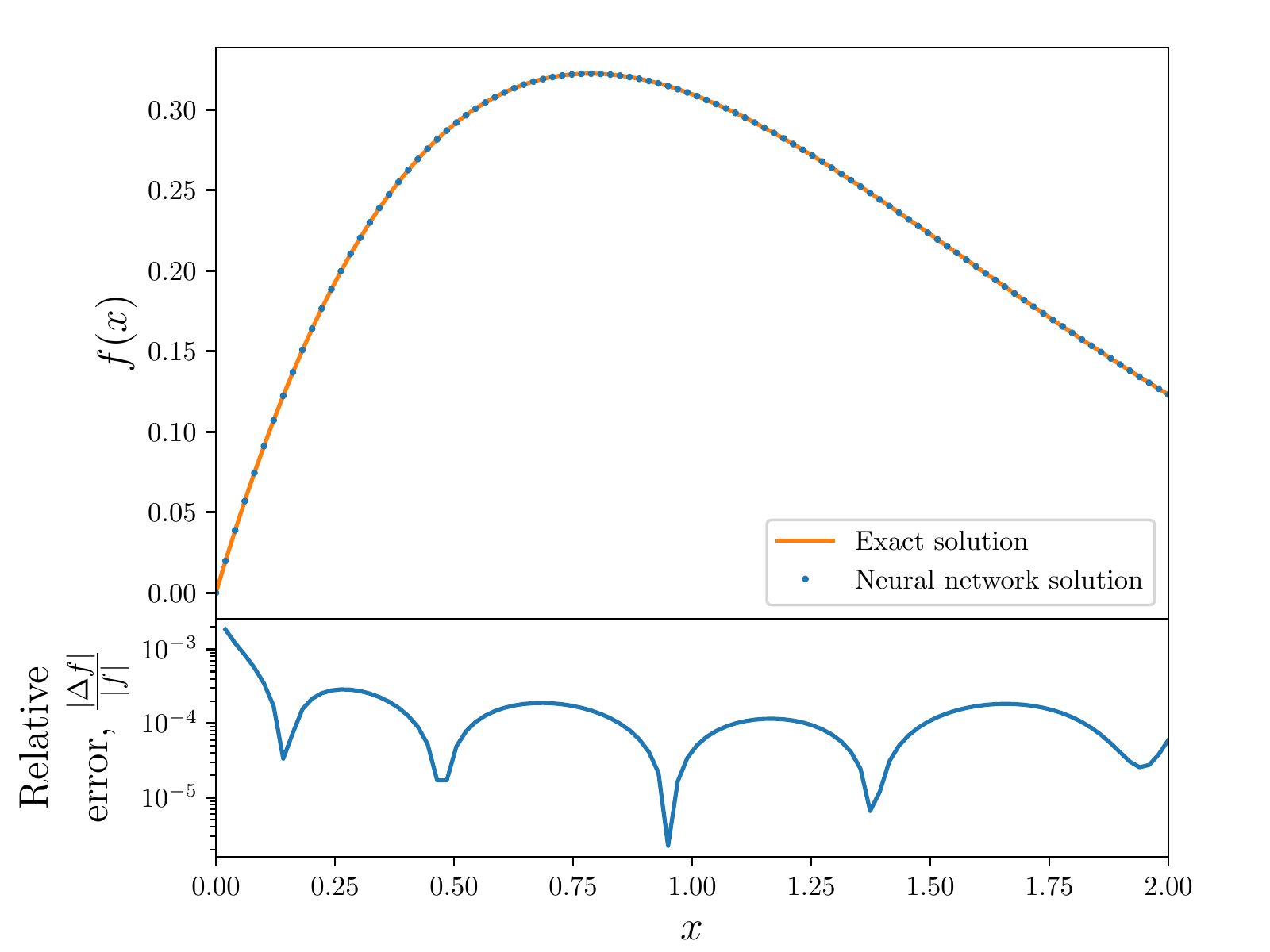}
	\caption{The NN solution (dots) of Eq. (\ref{first_order}) subject to the initial value $f(0) = 0$ overlaid on the exact solution (solid line). The lower panel shows the relative error. The network was trained for $10^4$ epochs using Adam optimizer with 100 uniformly spaced training points in [0,2]. One hidden layer of 10 neurons with a sigmoid activation function was used. }
	\label{fig:First_order_plot}
\end{figure}

The analytic solution is $f(x) = e^{-x}\sin(x)$. The equation being simple, the NN is expected to reach an acceptable accuracy for relatively few epochs. The accuracy and efficiency of the solver are illustrated in \textbf{FIG.} \ref{fig:First_order_plot} where the NN has been trained for $10^4$ epochs reaching $1.6\times 10^{-4}$ as mean relative error. The bottom panel in \textbf{FIG.} \ref{fig:First_order_plot} exhibits clear wavy patterns showing that the NN solution wiggles around the exact solution. Better accuracy can be found by increasing the number of epochs.

    \subsection{Stationary Schr\"odinger equation}
    \label{subsec:schro}
    
The one dimensional time-independent Schr\"odinger equation, with potential $V(x)$, has the following differential equation\footnote{We set $\hbar = 1$.}:

\begin{figure*}[htp]
	\centering
	\begin{minipage}{.5\textwidth}
		\centering
		\includegraphics[height=0.275\textheight]{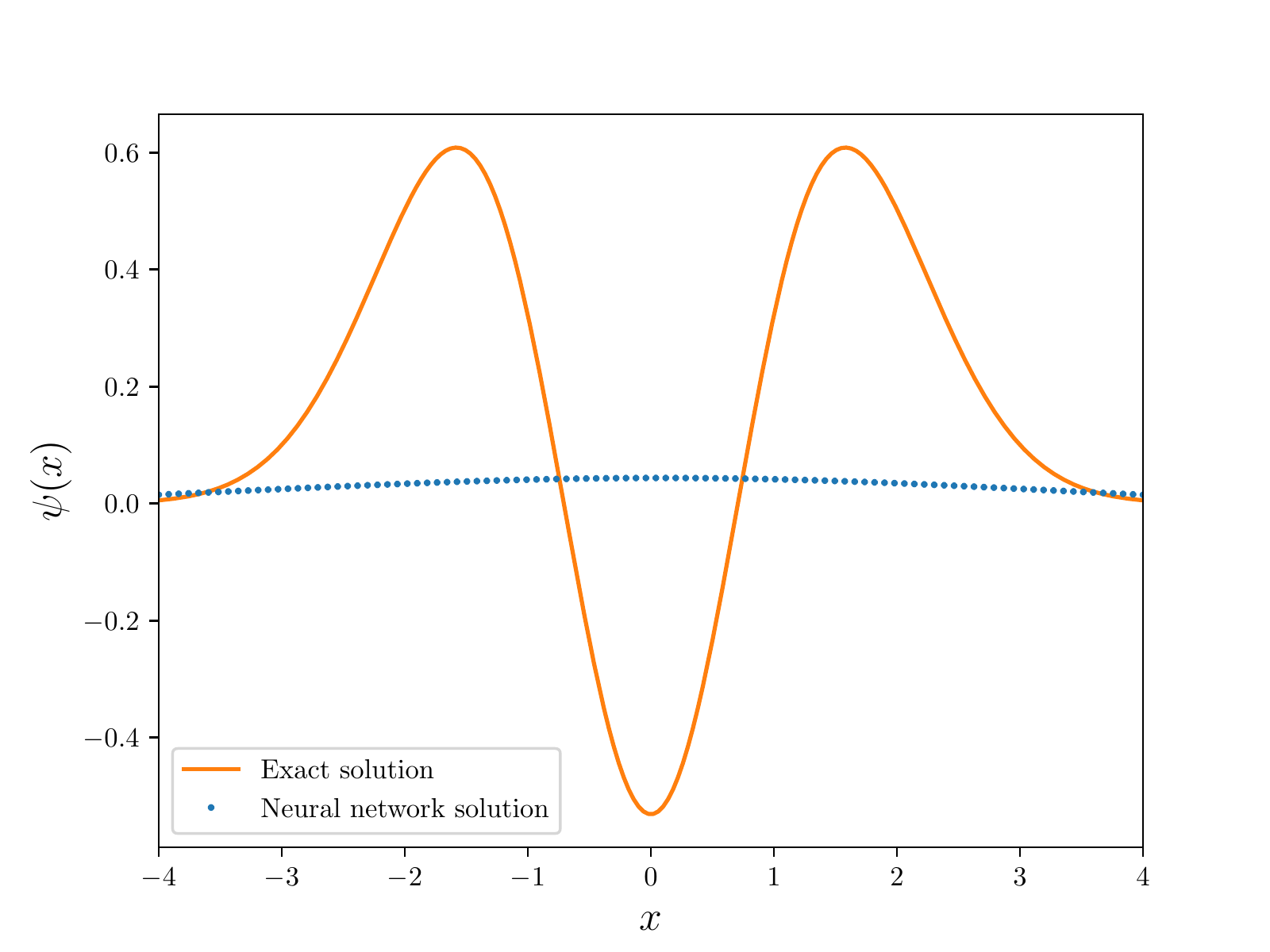}
		\subcaption{$10^3$ epochs}
		\label{fig:se_1000_epoch}
	\end{minipage}%
	\begin{minipage}{.5\textwidth}
		\centering
		\includegraphics[height=0.275\textheight]{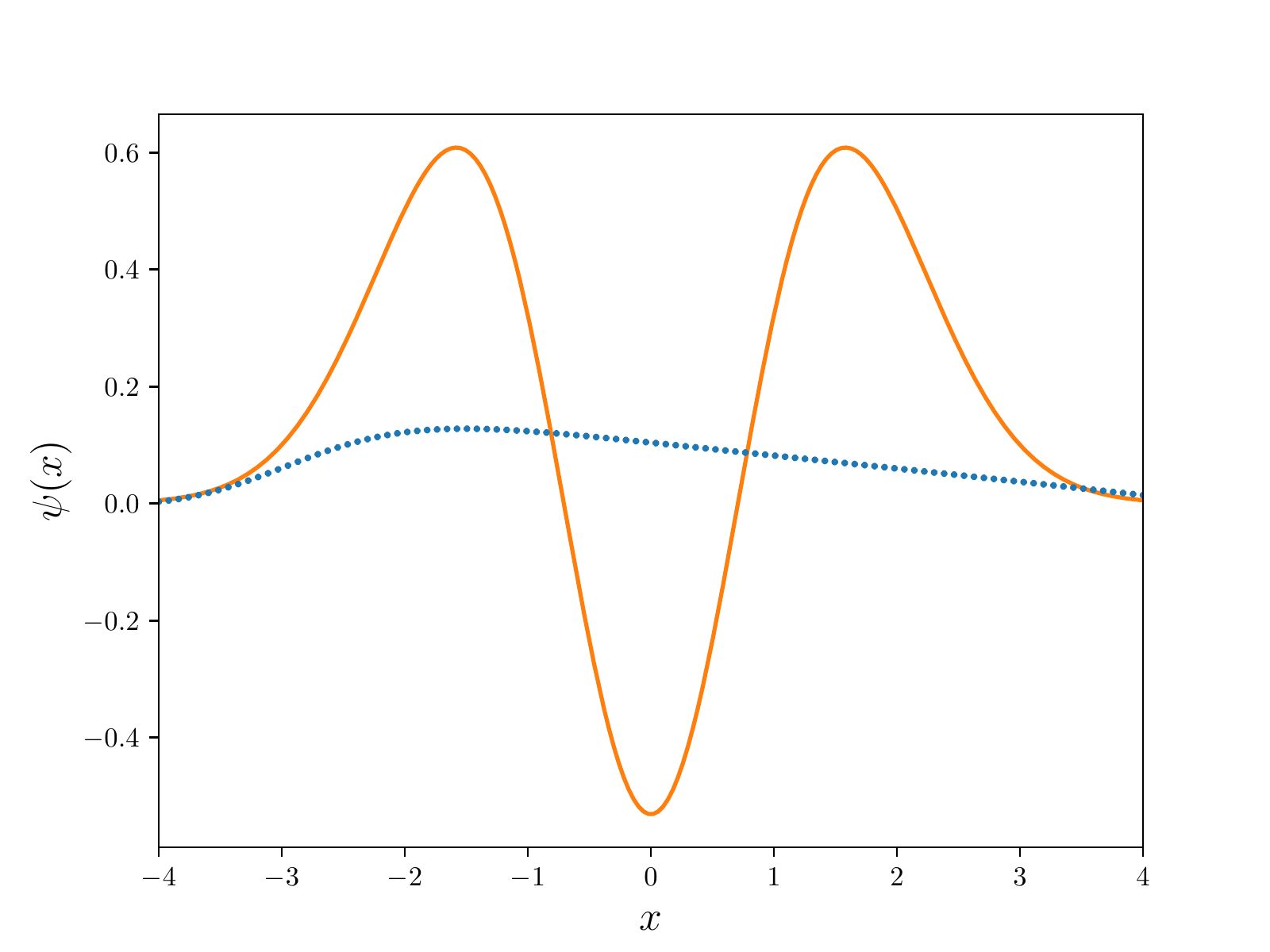}
		\subcaption{$6\times 10^3$ epochs}
		\label{fig:se_6000_epoch}
	\end{minipage}%
	
	\begin{minipage}{.5\textwidth}
		\centering
		\includegraphics[height=0.275\textheight]{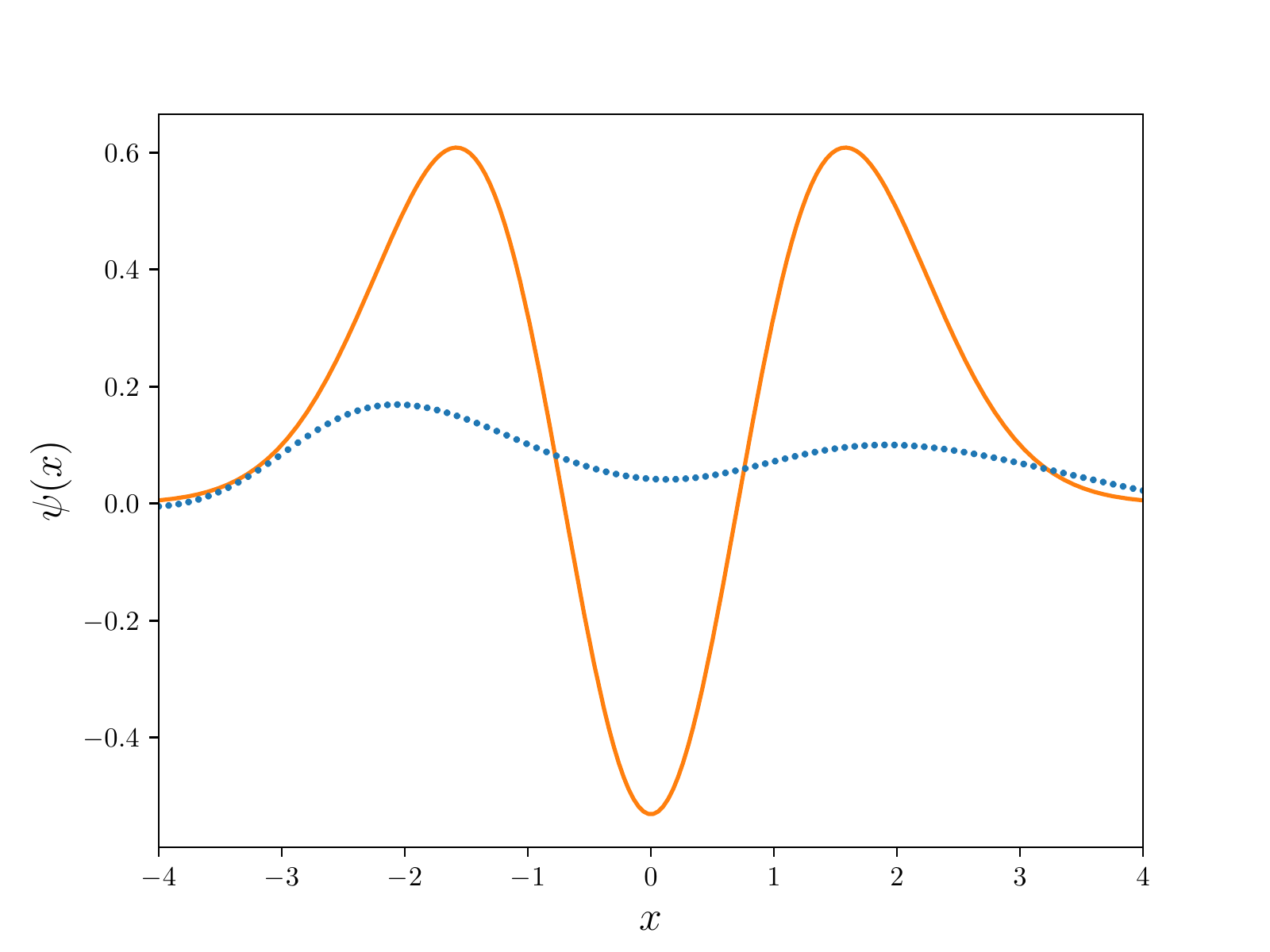}
		\subcaption{$7\times 10^3$ epochs}
		\label{fig:se_7000_epoch}
	\end{minipage}%
	\begin{minipage}{.5\textwidth}
		\centering
		\includegraphics[height=0.275\textheight]{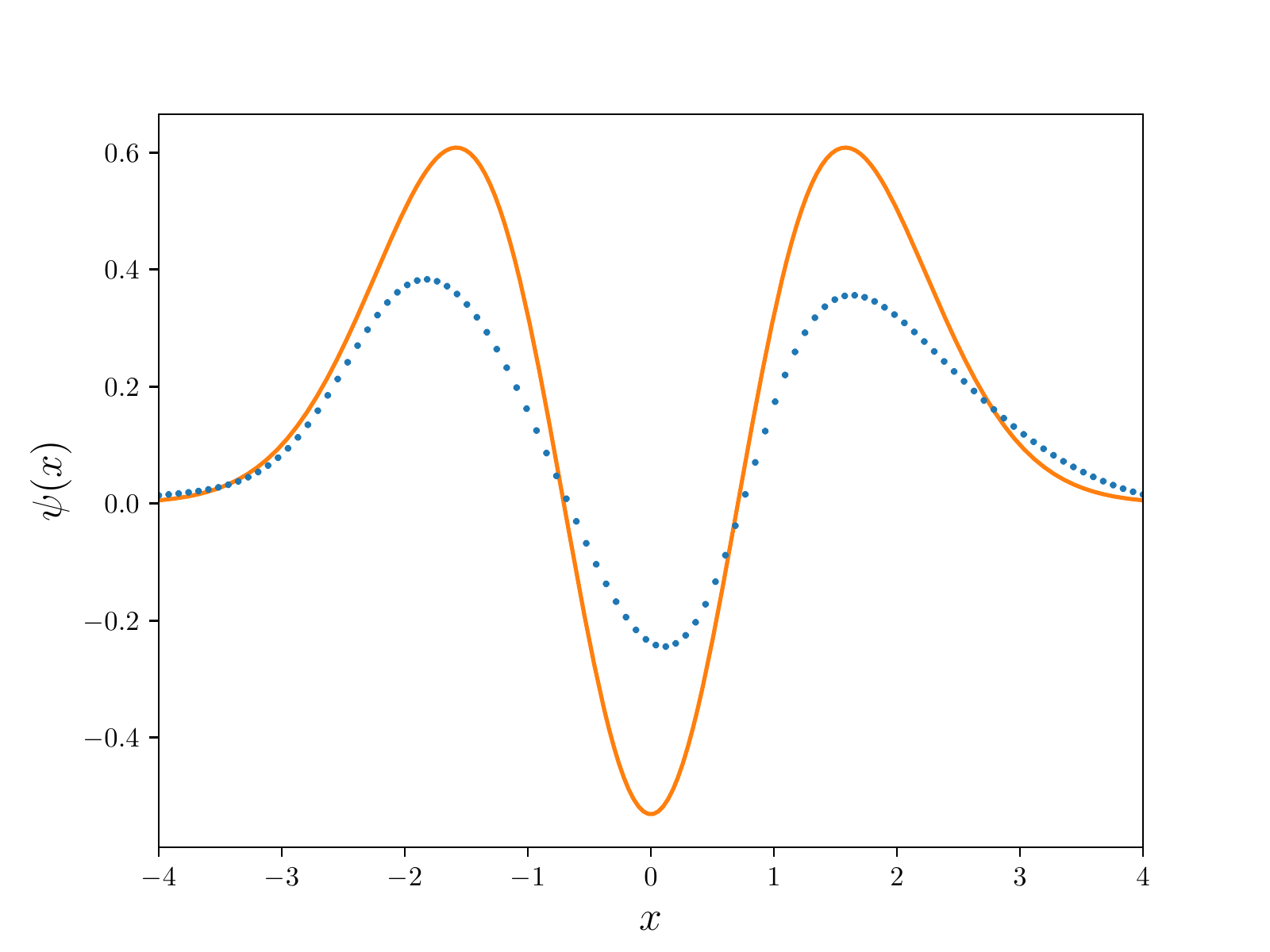}
		\subcaption{$8\times 10^3$ epochs}
		\label{fig:se_8000_epoch}
	\end{minipage}%
	
	\begin{minipage}{.5\textwidth}
		\centering
		\includegraphics[height=0.275\textheight]{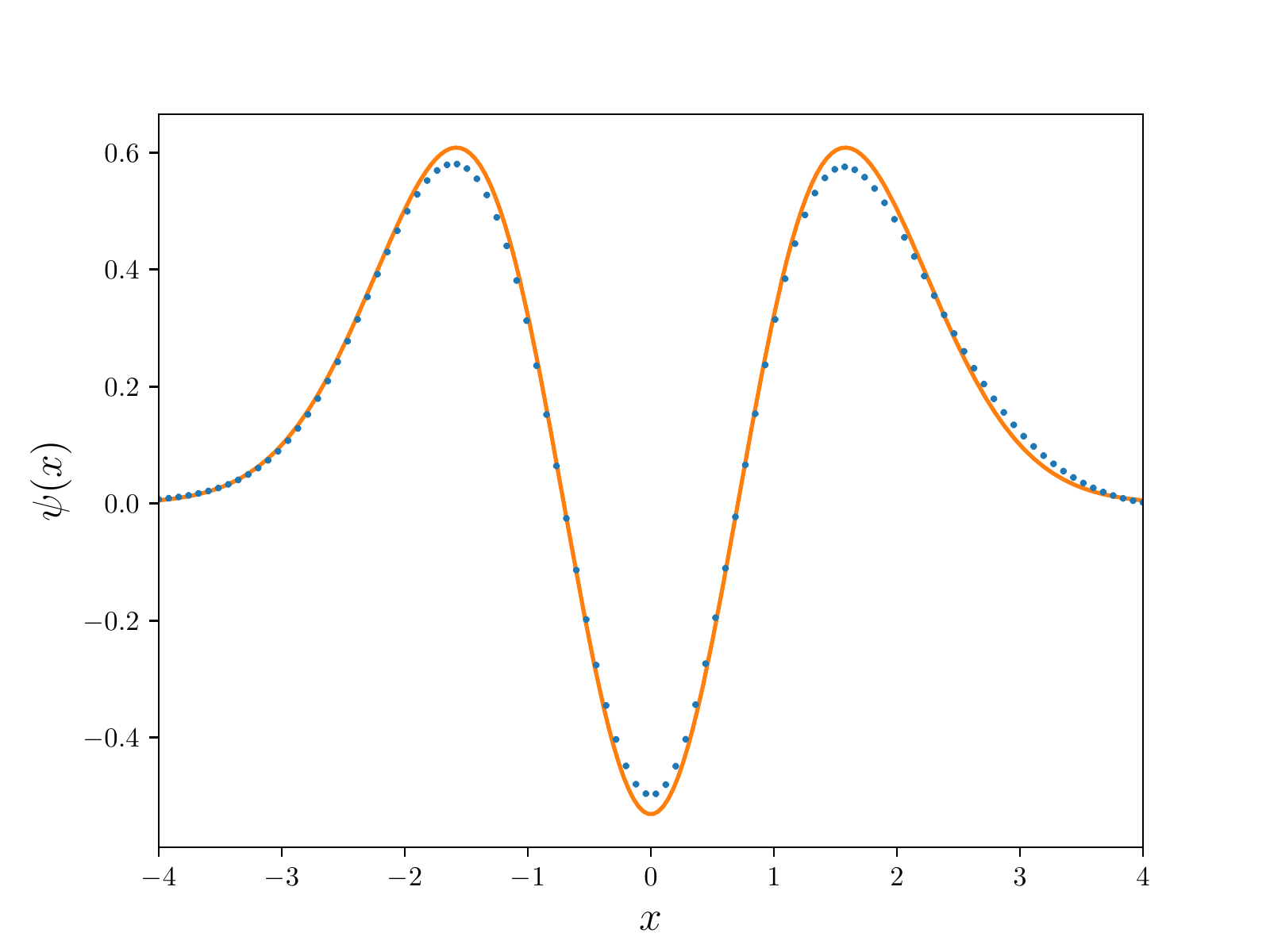}
		\subcaption{$10^4$ epochs}
		\label{fig:se_10000_epoch}
	\end{minipage}%
	\begin{minipage}{.5\textwidth}
		\centering
		\includegraphics[height=0.275\textheight]{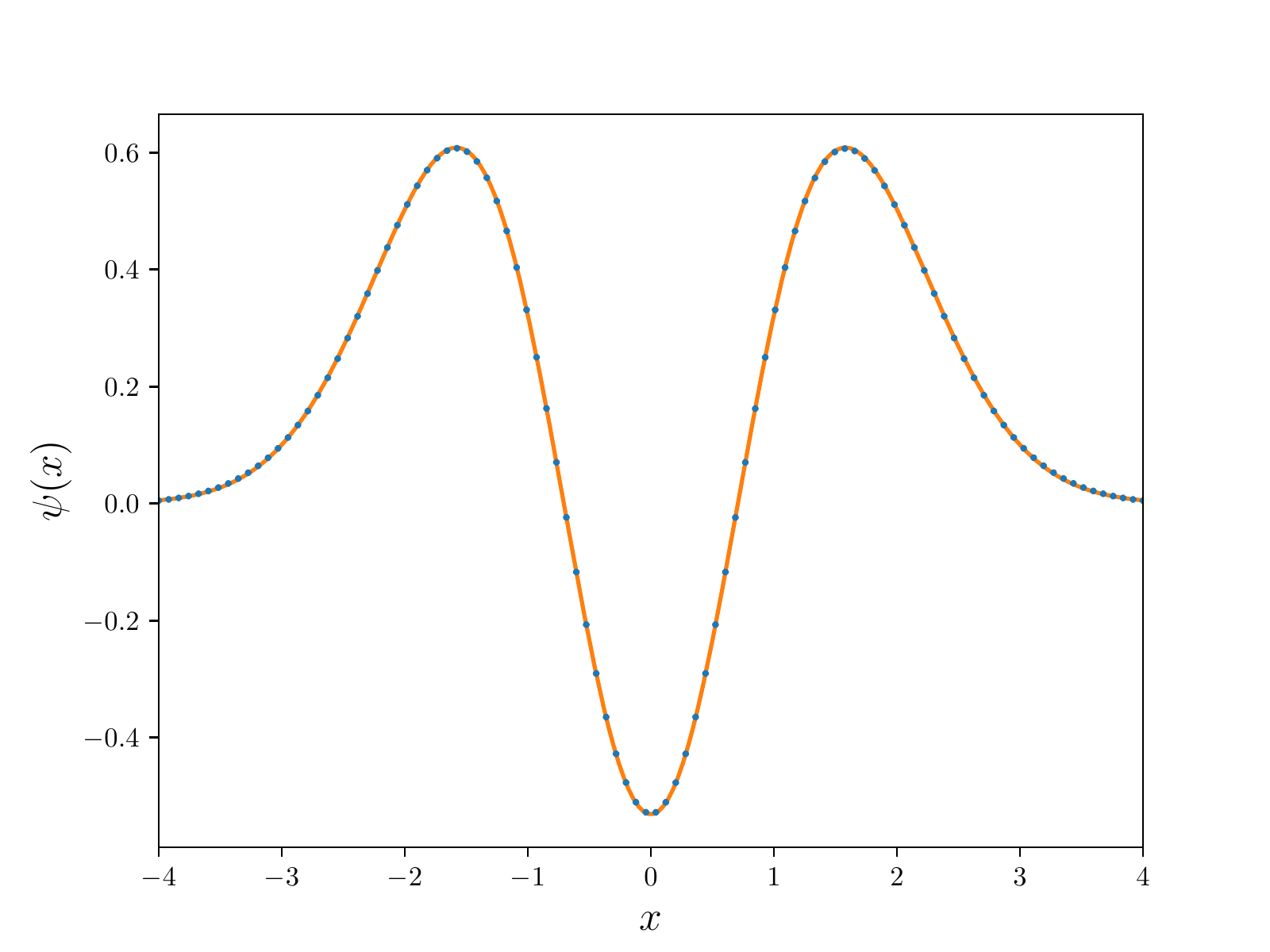}
		\subcaption{$1.8\times 10^4$ epochs}
		\label{fig:se_18000_epoch}
	\end{minipage}
	\caption{The NN $n = 2$ energy eigenfunction (dots) as a solution of Eq. (\ref{harmonic_se}) along with the exact solution (solid line), shown for comparative purposes, at different epochs during the training. The same boundary conditions, architecture, activation function, and domain sampling as \textbf{FIG.} \ref{fig:schrodinger_plot_n_1} were used.}
	\label{fig:se_n2_epochs}
\end{figure*}

\begin{equation}
\label{harmonic_se}
\frac{\mathrm{d}^2\psi}{\mathrm{d}x^2}(x) + 2m(E_n - V(x))\psi(x) = 0.
\end{equation}

\begin{figure}[h!]
	\centering
	\hspace*{-0.5cm}
	\includegraphics[width=0.5\textwidth]{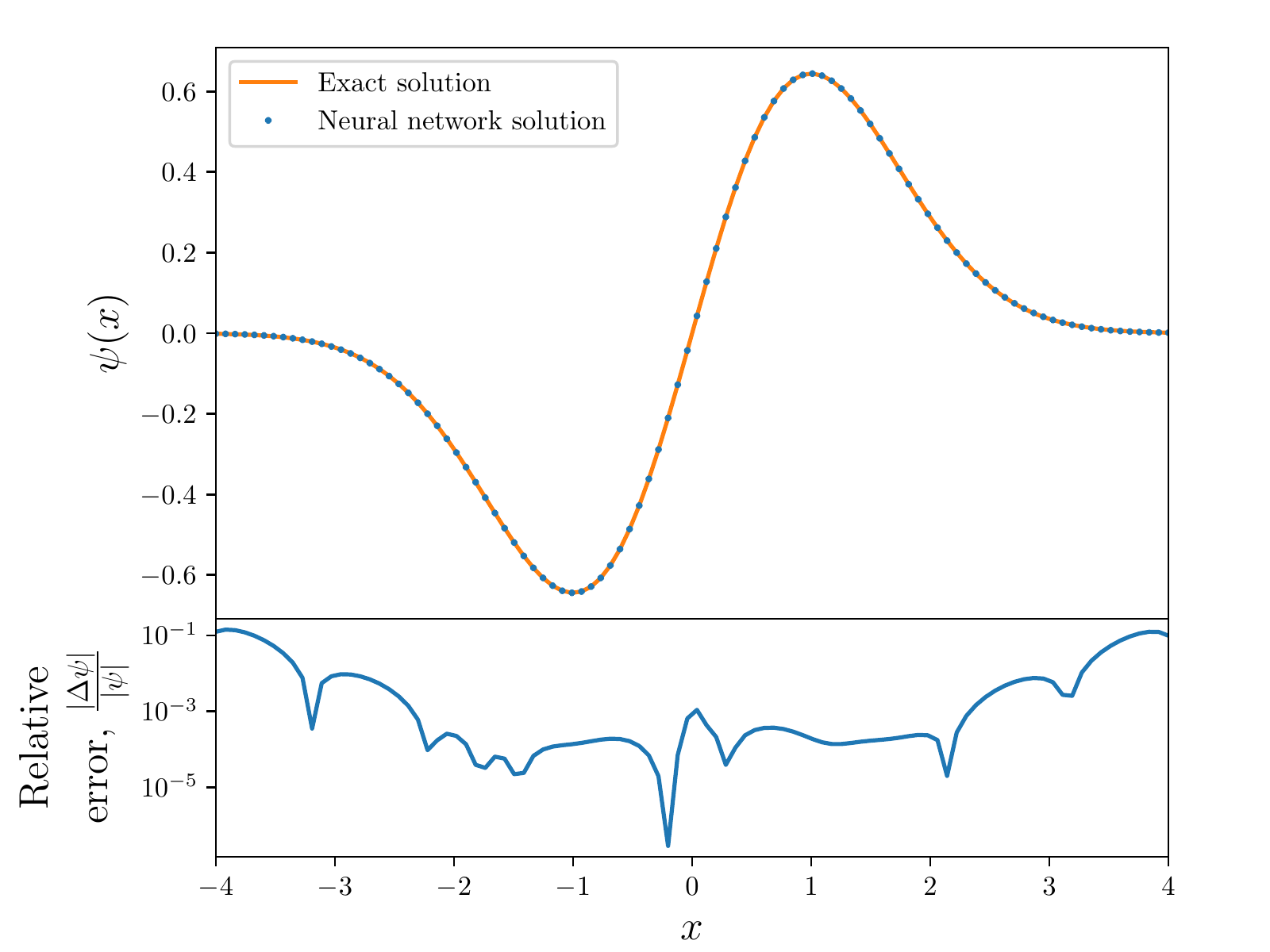}
	\caption{NN solution (dots) of Eq. (\ref{harmonic_se}) for $n = 1$ overlaid on the exact solution (solid line). Dirichlet boundary conditions are imposed at $x = \pm 2$. The lower plot shows the relative error. The network was trained for $5\times 10^4$ epochs using Adam optimizer with 100 uniformly spaced training points in $[-5,5]$ (but displayed for $[-4,4]$). One hidden layer of 50 neurons with a sigmoid activation function was used.}
	\label{fig:schrodinger_plot_n_1}
\end{figure}

The energy $E_n$ is quantized with an integer $n$. For a harmonic potential $V(x) = x^2$, the $n^{\text{th}}$ level has energy $n + 1/2$ with a corresponding analytical solution to the $n^{th}$ energy eigenfunction $\psi_n(x)$ given in terms of the Hermite polynomials. \textbf{FIG.} \ref{fig:schrodinger_plot_n_1} shows the NN evaluation of the energy eigenfunction $\psi(x)$ for $n = 1$. In this example, Dirichlet boundary conditions are imposed at $x = \pm 2$. Note that the relative error is maintained at $\sim 10^{-4}$ in between the boundary conditions and degrades outside. The NN is found to perform well on solving a boundary-value problem. As the exact solution rapidly vanishes outside the boundary conditions, the relative error dramatically increases so that it is no longer a valid measure of accuracy. 

The required number of epochs during the training can only be found by iterative testing. \textbf{FIG.} \ref{fig:se_n2_epochs} shows the neural network prediction for the $n = 2$ energy eigenfunction for $x\in[-4,4]$ at different epochs. One can see a general tendency in the training process that was found for other equations as well: (i) first, the NN fits the general trend of the solution by adjusting its prediction along a tangent of the curve (\textbf{FIGs.} \ref{fig:se_1000_epoch} and \ref{fig:se_6000_epoch}), (ii) then, the NN fits the curvature to reveal the true shape of the solution (\textbf{FIGs.} \ref{fig:se_7000_epoch}-\ref{fig:se_10000_epoch}), (iii) finally, it fine-tunes to adjust the prediction to the exact solution, decreasing the relative error (\textbf{FIG.} \ref{fig:se_18000_epoch}). These three phases of the training can be used to guess the number of epochs required to reach the desired relative error. Generally, the more complex the solution is, the more epochs are necessary to let the NN fit the curvature. Once the right curve is found, increasing the number of epochs increases the accuracy.

    \subsection{Burst equation}
    
To illustrate the NN's ability to fit a complex shape, we solve the following second-order differential equation
\begin{equation}
\label{burst}
\frac{\mathrm{d}^2f}{\mathrm{d}x^2}(x) + \frac{n^2 - 1}{(1+x^2)^2} f(x) = 0.
\end{equation}

The solution of Eq. (\ref{burst}) is characterized by a burst of approximately $n/2$ oscillations in the region $|x| < n$ \cite{Agocs}. An analytical solution for the equation is

\begin{equation}
\label{burst_exact}
f(x) = \frac{\sqrt{1+x^2}}{n} \cos(n \, \arctan \, x).
\end{equation}

\begin{figure}[h!]
	\centering
	\hspace*{-0.5cm}
	\includegraphics[width=0.5\textwidth]{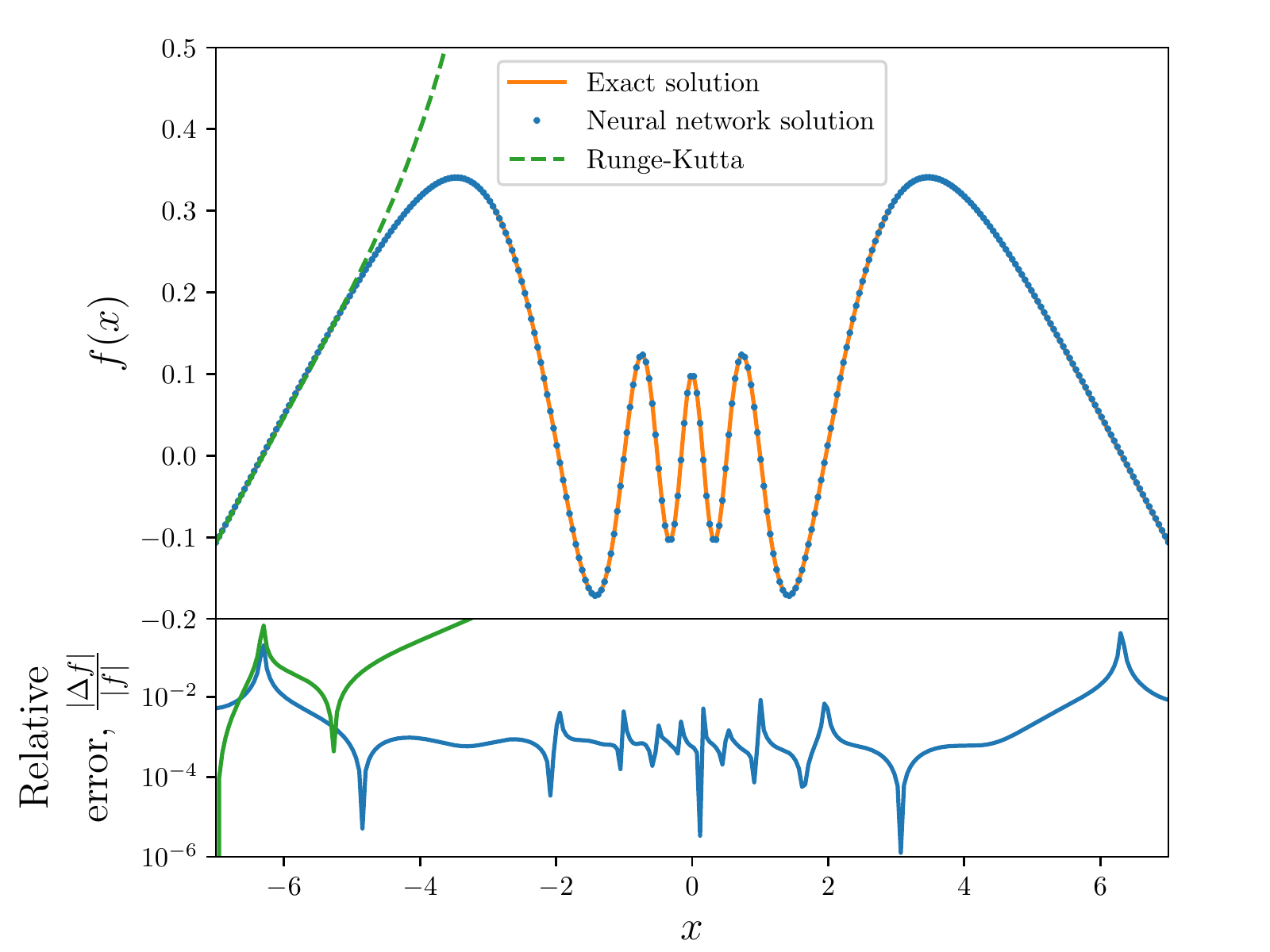}
	\caption{NN solution (dots) of Eq. (\ref{burst}) for $n = 10$ overlaid on the exact solution Eq. (\ref{burst_exact}) (solid line). Dirichlet boundary conditions are imposed at $x = 1.5$ and $x = 3$. The lower plot shows the relative error. The network was trained for $2\times 10^5$ epochs with minibatches of size 30 using Adamax optimizer with 300 uniformly spaced training points in $[-7,7]$. Three hidden layers, each having 30 neurons with a tanh activation function were used.}
	\label{fig:burst}
\end{figure}

The exact solution and the NN prediction is shown in \textbf{FIG.} \ref{fig:burst} for $n=10$. With usual numerical solvers such as Runge-Kutta based approaches, the relative error grows exponentially during the step by step integration, failing to reproduce the oscillations of Eq. (\ref{burst_exact}) (see \textbf{FIG.} \ref{fig:burst}) \cite{Agocs}. With an optimization-based approach, the trained NN is able to capture the behavior of the rapid oscillations, albeit requiring the training of a more complicated architecture over a greater number epochs when compared to the first order differential equation example in \textbf{FIG.} \ref{fig:First_order_plot} and the Schr\"odinger equation example in \textbf{FIG.} \ref{fig:schrodinger_plot_n_1}. Training on minibatches of the discretized domain was found to outperform training using the whole domain as a batch. Even for boundary conditions being imposed only at positive $x$ values, the network is able to retain the symmetry of the solution for negative $x$ values without losing accuracy.

\section{Performances}
\label{sec:perfomances}

We find that even if the NNs are not susceptible to overfitting within the training domain, the main issue of the NN approach arises from the arbitrariness in the choice of NN hyperparameters and the choice of the NN architecture. The flexibility of the method enables us to obtain a thorough understanding of how well different NN models perform on solving differential equations.

Traditional non-linear activation functions such as sigmoid and tanh are found to reproduce the ODE solution with less training than modern activation functions such as ReLU, ELU, and SELU. Rather than using a random uniform distribution or constant initialization, we find that the properly scaled normal distribution for weight and bias initialization called Xavier initialization\footnote{GlorotNormal in \textsf{Tensorflow 2}.} \cite{glorot} \cite{Zhang}, makes the training process converge faster. Xavier initialization is used for all figures. Also, Adam based approaches including Adamax and Nadam, with a learning rate set to $10^{-3}$, are found to perform better than stochastic gradient descent. 

    \subsection{Identifying the loss function with the mean squared error}
    \label{subsec:loss}

To evaluate the accuracy of the NN solution and the training performance, we suggest to use the loss function $\mathcal{L}$ in Eq. (\ref{cost_function}). Indeed, we show that the loss function as written in Eq. (\ref{cost_function}) can be identified with the mean absolute error computed with the $L^2$ norm, with some scaling. The identification makes the loss function an excellent indicator of how accurate the found solution is without the need of an exact solution. Furthermore, the numerical stability and the computational cost of the training process can be assessed without computing the relative error, as it is automatically encrypted in the loss function (\ref{cost_function}).

The main idea of this identification is to write the loss function (\ref{cost_function}) as a continuous functional of the unknown solution $f$ and Taylor expand it around an exact solution $f^0$ to the second order. Using the same notations as in Section \ref{sec:methodology}, let us take the continuous limit of the first term in Eq. (\ref{cost_function}), turning the sum into an integral, and defining the loss $\mathcal{L}$ as a functional $\mathcal{L}[f]$

\begin{equation}
\mathcal{L} \rightarrow \mathcal{L}[f],
\end{equation}
with

\begin{equation}
\label{functional}
\begin{aligned}
\mathcal{L}[f] &= \int_{\mathcal{D}}\mathrm{d}x\, \mathcal{F}[x, f(x), \nabla f(x), ..., \nabla^{p}f(x)]^2 \\
&+ \sum_{j} [\nabla^{k}f(x_j) - K_{j}]^2.
\end{aligned}
\end{equation}

The loss functional has to be minimized on the entire differential domain $\mathcal{D}$ rather than on a set of discrete points $x_i\in\mathcal{D}$. Using the physics language, $\mathcal{L}[f]$ is to be understood as the action with some Lagrangian $\mathcal{F}[x, f(x), \nabla f(x), ... \nabla^{p}f(x)]^2$. For clarity, we define $\bar{\mathcal{F}} = \mathcal{F}^2$. The Cauchy-Peano theorem guarantees the existence of an exact solution $f^0$ to differential equations subject to boundary/initial conditions. The solution $f^0$ minimizes the loss function making it vanish $\mathcal{L}[f^0] = 0$. We now Taylor expand $\mathcal{L}[f]$ around $f^0$ to the second order

\begin{equation}
\label{Taylor}
\begin{aligned}
&\mathcal{L}[f] = \int_{\mathcal{D}}\mathrm{d}x\, \bar{\mathcal{F}}[x, f^0(x), \nabla f^0(x), ..., \nabla^{p}f^0(x)] \\
&+ \int_{\mathcal{D}}\mathrm{d}x\, \frac{\delta \bar{\mathcal{F}}}{\delta f}[x, f^0(x), \nabla f^0(x), ...  ,\nabla^{p}f^0(x)]\, \left(f(x) - f^0(x)  \right) \\
&+ \int_{\mathcal{D}}\mathrm{d}x\, \frac{\delta^2 \bar{\mathcal{F}}}{\delta f^2}[x, f^0(x), \nabla f^0(x), ... , \nabla^{p}f^0(x)]\, \left(f(x) - f^0(x)  \right)^2 \\
&+ \mathcal{O}\left(\left[f(x) - f^0(x)  \right]^3\right).
\end{aligned}
\end{equation}
 
The second term in Eq. (\ref{functional}) vanishes because $f^0$ satisfies the boundary/initial conditions. The zero order term in Eq. (\ref{Taylor}) vanishes because $f^0$ is an exact solution \textit{i.e.} $\bar{\mathcal{F}}[x, f^0(x), \nabla f^0(x), ... \nabla^{p}f^0(x)] = 0$ for all $x\in\mathcal{D}$. The first order term in Eq. (\ref{Taylor}) vanishes because $f^0$ minimizes the loss function so that the functional derivative is zero\footnote{By analogy with physics, this term would be the Euler-Lagrange equation after an integration by parts.} \textit{i.e.} $\frac{\delta \bar{\mathcal{F}}}{\delta f}[x, f^0(x), \nabla f^0(x), ... \nabla^{p}f^0(x)] = 0$ for all $x\in\mathcal{D}$. The remaining non-vanishing term is the second order term. As the $\frac{\delta^2 \bar{\mathcal{F}}}{\delta f^2}[x, f^0(x), \nabla f^0(x), ... \nabla^{p}f^0(x)]$ term does not depend on the NN solution and so does not depend on the parameters $\bm{\theta}$, it is a function of $x$ only and can be seen as a scaling $\alpha(x)$

\begin{equation}
\mathcal{L}[f] =  \int_{\mathcal{D}}\mathrm{d}x\, \alpha(x)\, \left[f(x) - f^0(x)  \right]^2 \\
+ \mathcal{O}\left(\left[f(x) - f^0(x)  \right]^3\right).
\end{equation}

Written in this form, the loss function appears to be the mean absolute error using the $L^2$ norm integrated over the entire domain $\mathcal{D}$, with some scaling function $\alpha(x)$. After discretizing $\mathcal{D}$ and identifying $f(x)$ with the neural network $\mathcal{N}(x, \bm{\theta})$, the loss function (\ref{cost_function}) takes the final form

\begin{equation}
\begin{aligned}
\mathcal{L}(\bm{\theta}) &\approx \frac{1}{N}\sum_{i=1}^N \alpha(x_i) \left[\mathcal{N}(x_i, \bm{\theta}) - f^0(x_i)  \right]^2\\
&+ \sum_{j} [\nabla^{k}\mathcal{N}(x_{j}, \bm{\theta}) - K_{j}]^2.
\end{aligned}
\end{equation}

Note that taking the continuous limit requires enough discrete points to make it valid. Here, NNs are usually trained with $\sim \mathcal{O}(100)$ points within the considered domains. Through this identification, we better understand the three training phases shown in section \ref{subsec:schro}. Indeed, as the loss function (\ref{cost_function}) can be Taylor expanded around the exact solution, the NN first minimizes the loss leading order term making it learn the general tendency of the solution first by finding the tangents. Then, the NN minimizes higher order terms, making the NN learn specific local patterns, like curvature. Finally, the NN adjusts the solution by considering even higher orders of the loss. These universal phases enable the method to converge locally with a stable accuracy that does not decrease over the entire training domain.

    \subsection{Domain sampling}

Our hypothesis is that using fine sampling for regions with more features, such as oscillations, would lead to greater accuracy \textit{i.e.} lower loss. Moreover, the computational cost of training the NN could be reduced by avoiding dense sampling of featureless regions in the domain. 

\begin{figure}[h!]
	\centering
	\hspace*{-0.5cm}
	\includegraphics[width=0.48\textwidth]{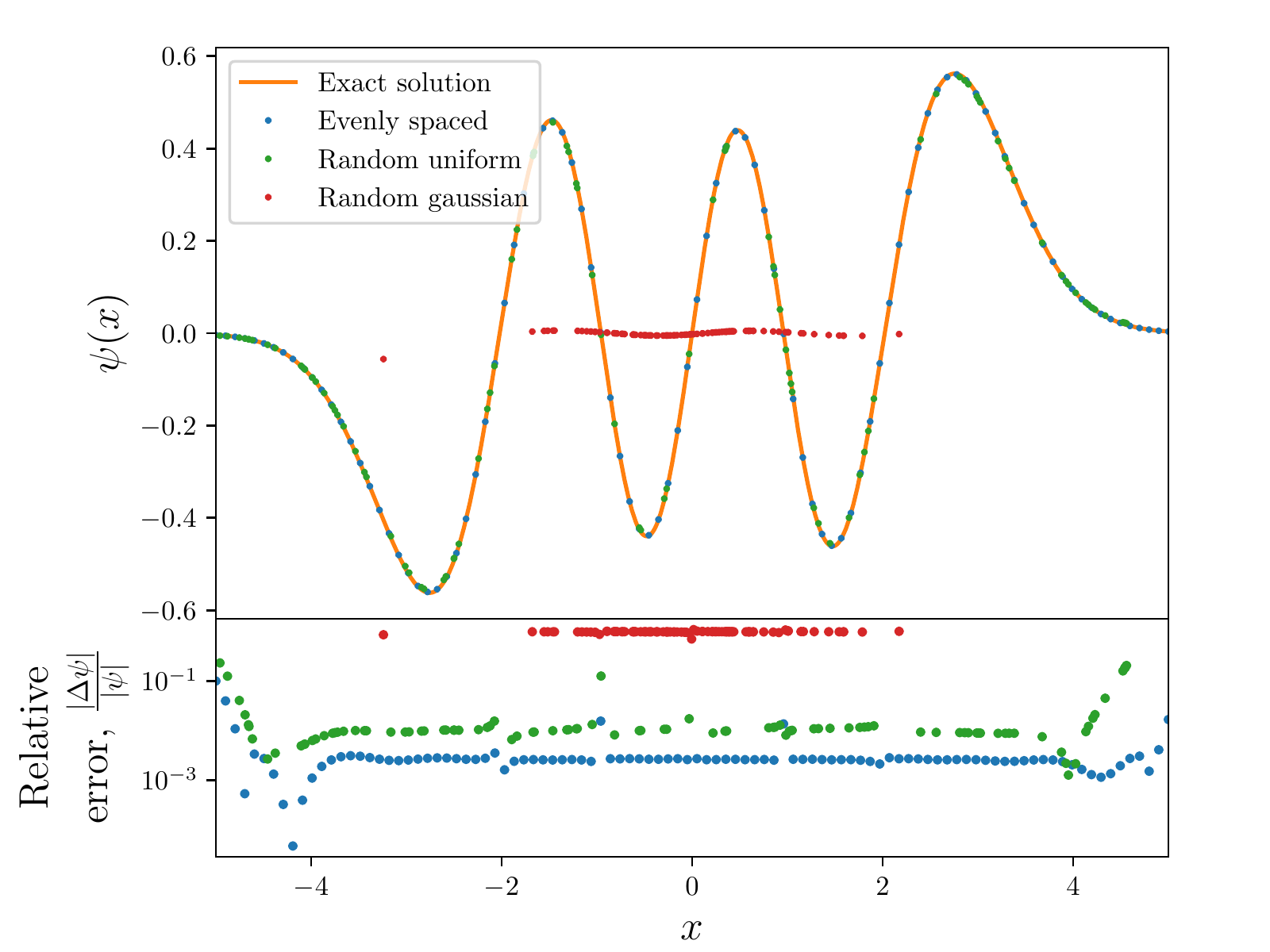}
	\caption{NN solution of Eq. (\ref{harmonic_se}) for $n=5$ overlaid on the exact solution (solid line) for (i) an evenly spaced training domain (blue dots), (ii) a random uniform training domain (green dots), and (iii) a random Gaussian training domain centered at $x=0$ with width 1 (red dots). The lower panel shows the relative error. All training domains contain 100 points in $[-5, 5]$. Dirichlet boundary conditions at $x=\pm 4$ were imposed. The network was trained for $4\times 10^5$ epochs using Adamax optimizer and tanh activation, and two hidden layers with 20 neurons in each layer were used.}
	\label{fig:sampling}
\end{figure}

We probe different domain samplings and expose their impact on the NN performance. \textbf{FIG.} \ref{fig:sampling} illustrates the effect of different domain samplings for the $n=5$ energy eigenfunction of Eq. (\ref{harmonic_se}): evenly spaced points, random uniformly distributed points, and points sampled from a Gaussian distribution centered at $x=0$ with standard deviation of 1, to accentuate the oscillatory region.

Surprisingly, a NN trained on evenly spaced points performs better than the other distributions. Every point in the region contributes equally in the loss function (\ref{cost_function}) and the NN learns the ODE solution uniformly over the entire domain. The uniform behavior can be seen in the lower panel of \textbf{FIG.} \ref{fig:sampling} as the relative error remains constant over the entire domain. Moreover, increasing the number of training points is found to have no influence on the NN performances but requires more training to reach the same accuracy. Randomizing the points is found to increase the relative error by an order of magnitude. Choosing a sampling distribution that accentuates a certain region fails to reproduce the ODE solution's shape for the same training time. Further training is then necessary to reproduce the same accuracy. 

\begin{figure}[ht!]
	\centering
	\hspace*{-0.5cm}
	\includegraphics[width=0.48\textwidth]{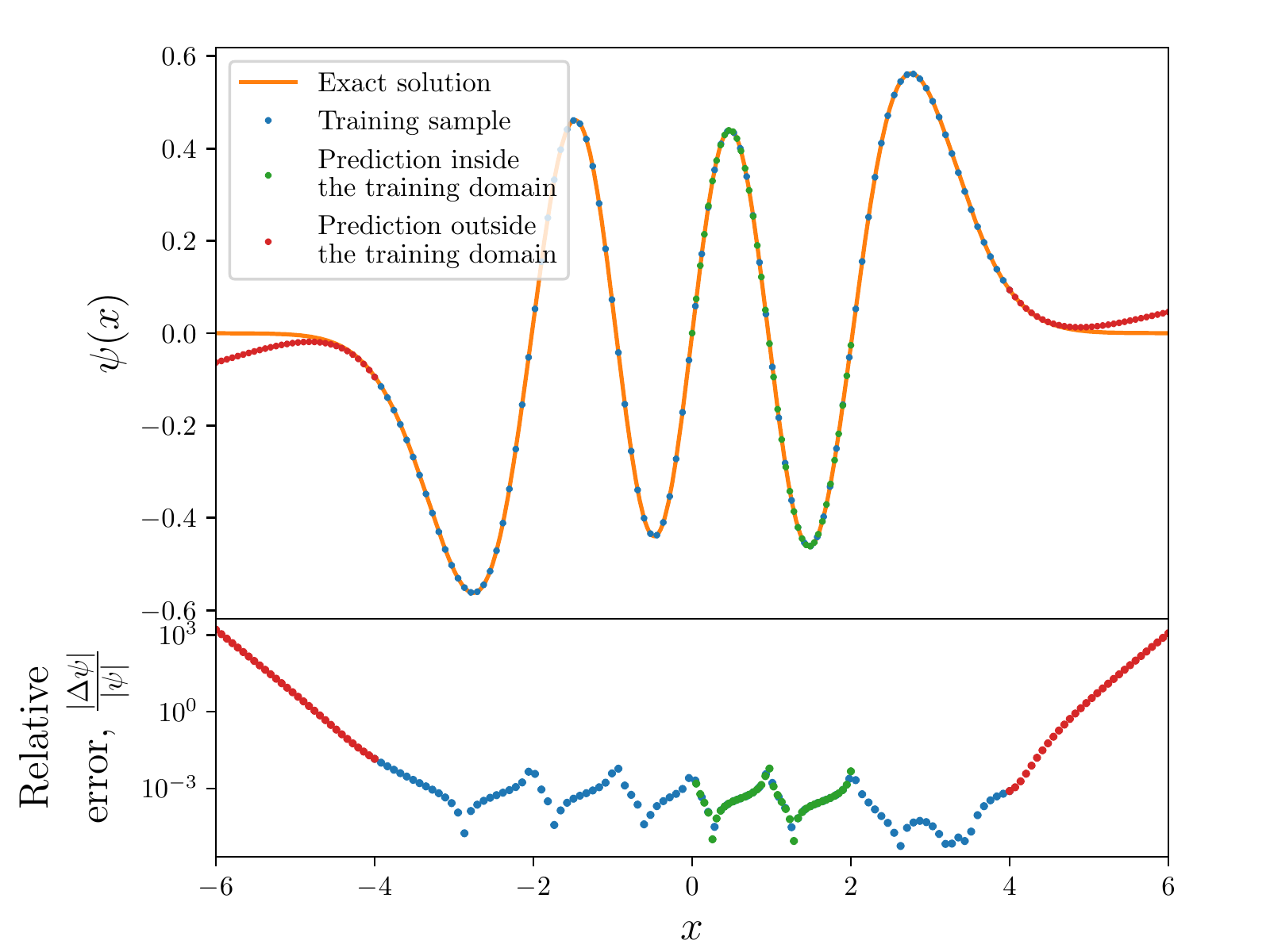}
	\caption{NN prediction of Eq. (\ref{harmonic_se}) for $n=5$ overlaid on the exact solution (solid line) for (i) the training points, 100 evenly spaced points in $[-4, 4]$ (blue dots), (ii) 40 evenly spaced points in $[0, 2]$, and (iii) 30 evenly spaced points in $[-6, -4]$ and $[4, 6]$. The lower panel shows the relative error. Dirichlet boundary conditions were imposed at $x = \pm 3$. The network was trained for $5\times 10^5$ epochs using three hidden layers with 30 neurons in each layer. The same optimizer and activation as in \textbf{FIG.} \ref{fig:sampling} were used.}
	\label{fig:extrapolation}
\end{figure}

    \subsection{Extrapolation performances}

Once the NN is trained, it is expected to be a continuous function within the finite, bounded training domain. 

Here, we illustrate the continuous nature of the NN solution. \textbf{FIG.} \ref{fig:extrapolation} shows the extrapolation performances for points inside the training domain and outside the training domain. The $n=5$ energy eigenfunction of Eq. (\ref{harmonic_se}) is solved within $[-4, 4]$ using 100 evenly spaced points. The NN prediction is then evaluated in $[0, 2]$, and in $[-6, -4]\cup [4, 6]$. 

The NN is able to reproduce the ODE solution with the same accuracy within the training domain but fails to reproduce the solution outside the training domain, even with more training. Note that the mean relative error is smaller than in \textbf{FIG.} \ref{fig:sampling} as more epochs were used for the training. In classical numerical analysis, the sample spacing \textit{i.e.} the integration step is usually small to avoid numerical instabilities \cite{beyond}. Here, the optimization approach enables us to train the NN on a reduced number of points, hence reducing the training time, and allows then to evaluate the prediction on a fine mesh.

    \subsection{Neural network architectures}

Choosing the exact architecture for a NN remains an art that requires extensive numerical experimentation and intuition, and is often times problem-specific. Both the number of hidden layers and the number of neurons in each layer can affect the performance of the NN. 

\begin{figure*}[hp!]
\centering
	\begin{minipage}{\textwidth}
		\centering
		\includegraphics[height=0.275\textheight]{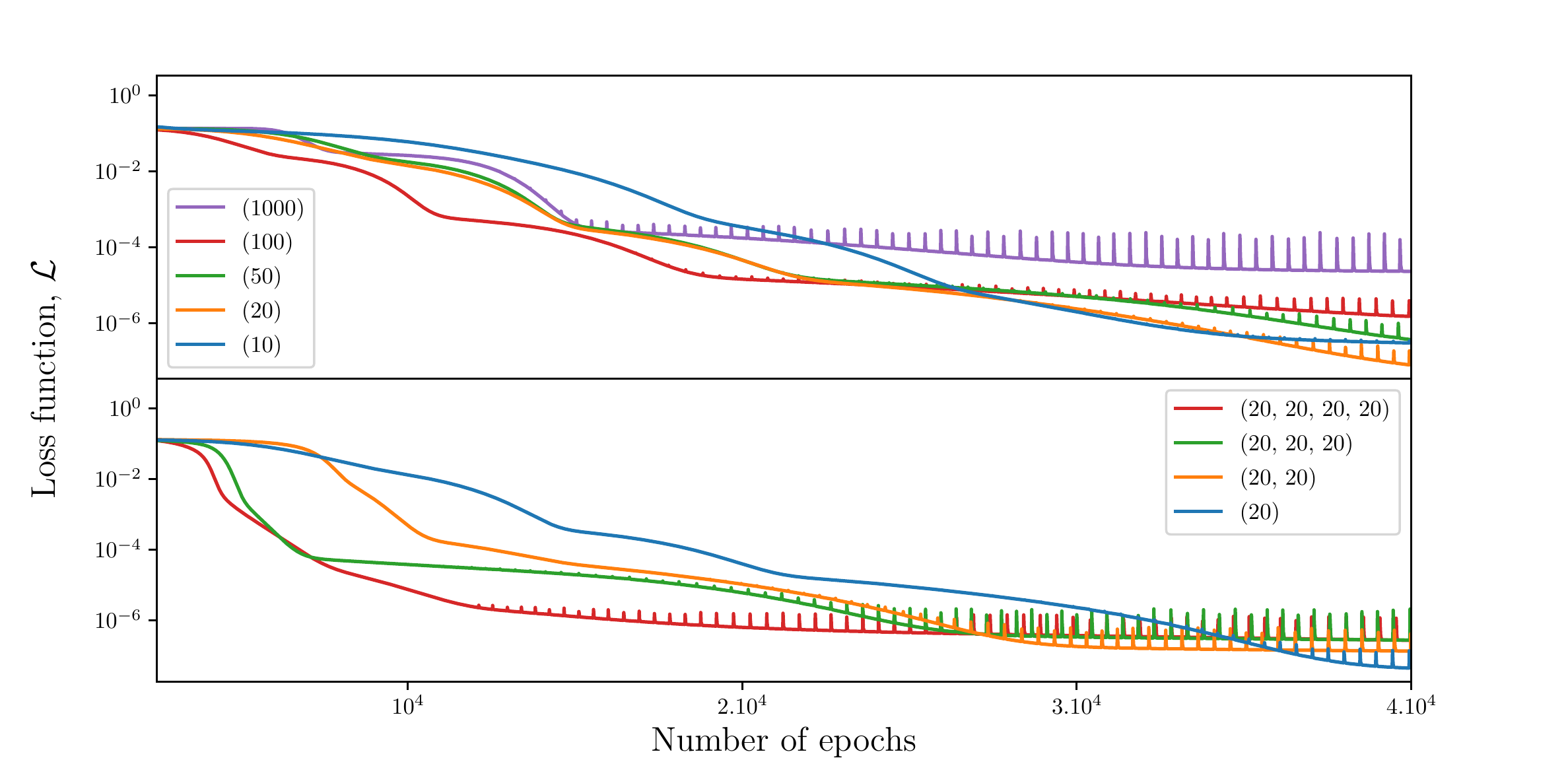}
		\subcaption{First order differential equation (\ref{first_order})}
		\label{fig:ode_arch}
	\end{minipage}%
	
	\begin{minipage}{\textwidth}
		\centering
		\includegraphics[height=0.275\textheight]{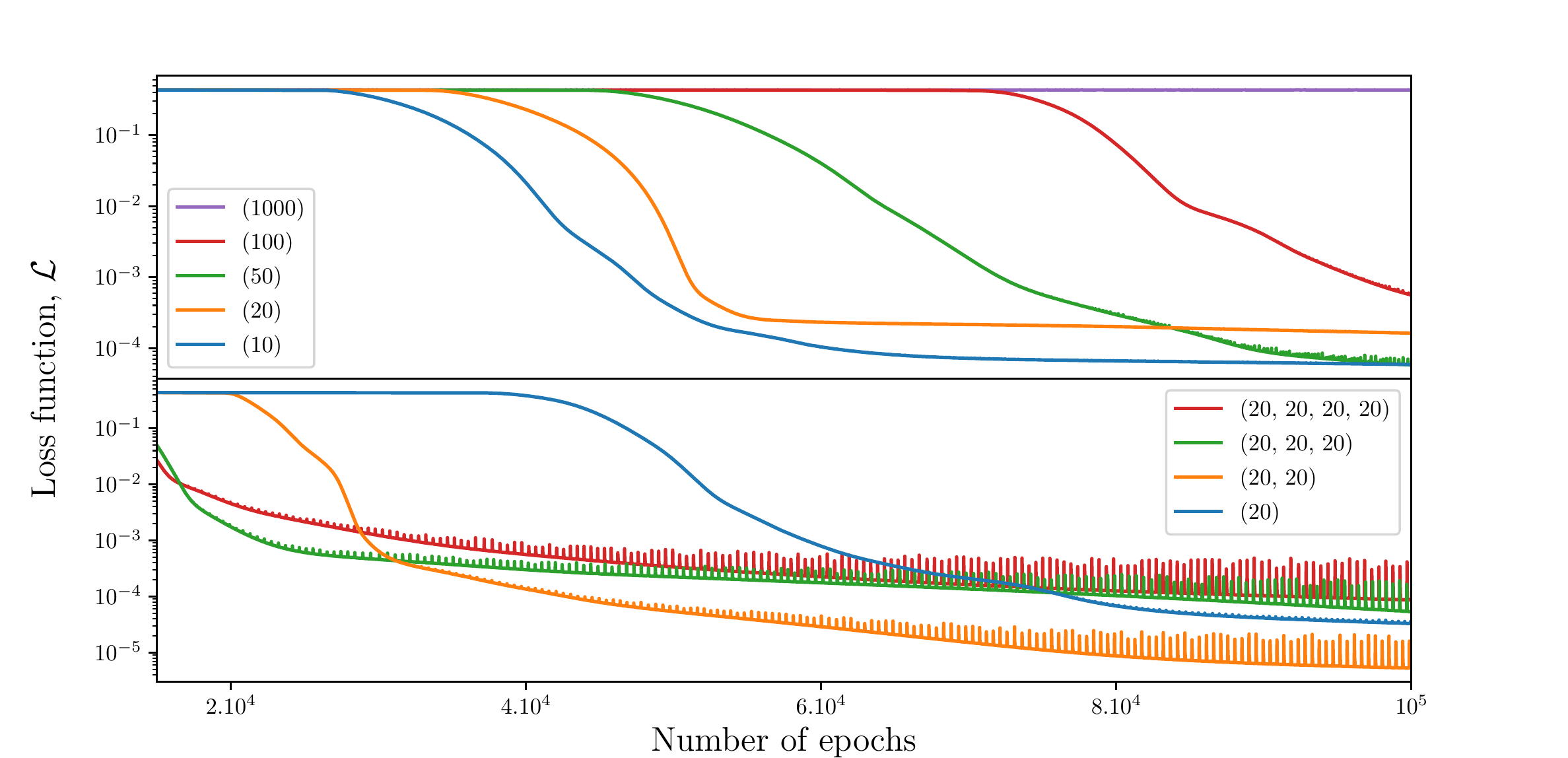}
		\subcaption{Schr\"odinger equation (\ref{harmonic_se}) for $n=2$}
		\label{fig:schro_arch}
	\end{minipage}%
	
	\begin{minipage}{\textwidth}
		\centering
		\includegraphics[height=0.275\textheight]{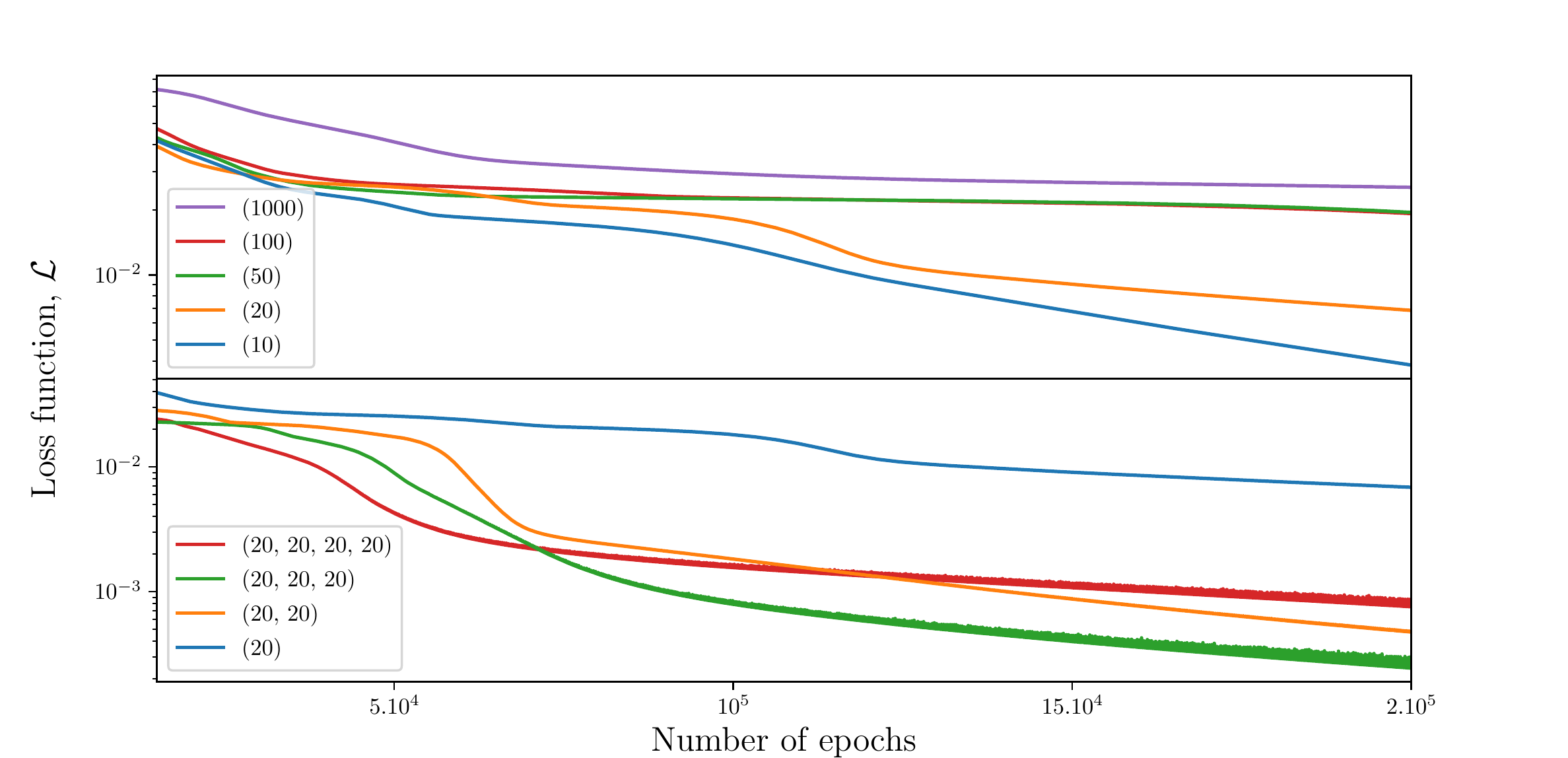}
		\subcaption{Burst equation (\ref{burst}) for $n = 10$}
		\label{fig:burst_arch}
	\end{minipage}%
	\caption{Loss function $\mathcal{L}$, proved to be the scaled mean squared error, during the training for different NN architectures. The first order ODE, the Schr\"odinger equation for $n=2$ and the Burst equation for $n=10$ are shown in \ref{fig:ode_arch}, \ref{fig:schro_arch} and \ref{fig:burst_arch} respectively. For each Figure, the upper panel is generated by varying the number of neurons in a single hidden layer, and the lower panel is generated by varying the number of hidden layers with 20 neurons in each layer. The same optimizers, activation functions and training domains as in \textbf{FIG.} \ref{fig:First_order_plot}, \ref{fig:schrodinger_plot_n_1} and \ref{fig:burst} for each equation were used. The learning rate was set $10^{-4}$ for each optimizer.}
	\label{fig:architecture}
\end{figure*}

A detailed systematic study to quantify the effect of the NN architecture in presented in \textbf{FIG.} \ref{fig:architecture}. We vary both the number of neurons in a single hidden layer and the number of hidden layers for all three equations (\ref{first_order}), (\ref{harmonic_se}) and (\ref{burst}). An unsatisfactory value of the loss after plateauing can be viewed as a poor NN architecture performance, as further training in general does not reduce its value. Here, the loss function is used to evaluate the NN performances during the training process as it can be identified with the scaled mean squared error (see section \ref{subsec:loss}).

Intuitively, one would expect an increasing NN complexity to enhance the accuracy. We found that some NN architectures, not necessary the more complex ones, perform better than others in reaching the desired accuracy for a certain number of epochs. In general, increasing the number of hidden layers make the loss decrease more rapidly in the beginning of the training until roughly $10^4$ epochs, but some simpler NN architectures prove to reach a better accuracy after more training. 

A clear link between the ODE complexity and the number of NN parameters is shown. The first order ODE (\ref{first_order}) being simple, a NN with one hidden layer containing 20 neurons is able to outperform a more complex NN with one hidden layer containing 1000 neurons (\textbf{FIG.} \ref{fig:ode_arch}). The difference affects the loss value by a few orders of magnitude. In this case, the number of hidden layers does not show any significant enhancement for the same amount of training. The Schr\"odinger equation (\ref{harmonic_se}) being more complex because it is a second order ODE, a two hidden layer NN performs better than a single hidden layer NN. However, the loss function for more than two hidden layers rapidly reaches a plateau and does not decrease after more training, making too complex NNs inefficient (\textbf{FIG.} \ref{fig:schro_arch}). Increasing the number of hidden layers for the Burst equation (\ref{burst}) to three is shown to both decrease the loss function faster and reach a smaller value than simple NNs (\textbf{FIG.} \ref{fig:burst_arch}). This tendency reveals the ODE complexity. 

Indeed, for all three equations, an empiric optimal NN architecture can be found and its complexity increases with the ODE complexity. The main limitation is that we find no clear recipe of how to guess the optimal architecture, as it has to be done by testing. However, we suggest that too complex NNs (more than three hidden layers and more than 100 neurons in each layer) should not be used as the problem of solving ODEs is in general too simple.

\vspace*{-0.15cm}
\section{Conclusion}
\label{sec:conclusion}
\vspace*{-0.1cm}
We have explored and critiqued the method of using NNs to find solutions to differential equations provided by Piscopo et al. \cite{Piscopo}. We found that the NN was able to perform better than typical numerical solvers for some differential equations, such as for highly oscillating solutions. We must note that the proposed method should not be seen as a replacement of numerical solvers as, in most cases, such methods meet the stability and performance required in practice.\\
\indent{}Our main message here is that the NN approach no longer appears as a black-box but a rather intuitive way of constructing accurate solutions. This approach was studied in detail. Three phases of fitting were characterized: first finding the general trend, secondly adjusting the curvature of the solution, and finally making small adjustments to improve the accuracy of the solution. Within the training domain, the NN was found to provide a continuous approximate function that matches the analytical solution to arbitrary accuracy depending on training time. However, extrapolation outside the training domain fails. We found that training in smaller minibatches rather than the whole discretized domain used in Lagaris et al. \cite{Lagaris} and Piscopo et al. \cite{Piscopo} gives a greatly reduced loss.\\
\indent{}A specifically designed loss function from the literature \cite{Piscopo} was proved to be the appropriate metric for evaluating the solution accuracy and the NN performances without the need of the exact solution, which is usually not known. Finally, we found the limitation of the method is finding a suitable architecture. There is no trivial relationship between the NN architecture and the accuracy of the NN approximate solution for a general differential equation, though a general tendency to increase the number of NN parameters to solve more complex differential equations was highlighted.\\
\indent{}A range of questions can be immediately explored. The performances of more sophisticated NN structures with dropouts and recurrent loops can be studied. Other sampling schemes can also be tested. Another question is whether the convergence of the NN solution to a certain accuracy can be achieved with fewer epochs. With the method we described, convergence comes locally, similar to a Taylor series. One might be able to reformulate the NN such that convergence comes globally, via a Fourier series representation, or using a different complete basis. Such reformulations might help the NN to learn general representations about the ODEs \cite{magill}. Global convergence may give better extrapolation results. Nevertheless, NNs show great potential as a support for standard numerical techniques to solve differential equations.

\vspace*{-0.1cm}
\begin{acknowledgements}
D.W. thanks the ENS Paris-Saclay for its continuing support via the normalien civil servant grant.
\end{acknowledgements}


\bibliography{bibliography}

\end{document}